\pdfoutput=1 
\documentclass[twocolumn,showpacs,prb,citeautoscript,amsmath,amssymb,floatfix]{revtex4}
\usepackage{dcolumn}
\usepackage{amsmath,amssymb}
\usepackage{bm}
\usepackage{graphicx}
\newcommand{\bd}{\bm}
\begin{document}

\title{Parametric pumping and kinetics of 
magnons in dipolar ferromagnets} 

\author{Thomas Kloss, Andreas Kreisel, 
and Peter Kopietz}
  
\affiliation{Institut f\"{u}r Theoretische Physik, Universit\"{a}t
  Frankfurt,  Max-von-Laue Strasse 1, 60438 Frankfurt, Germany}

\date{April 8, 2010}

 \begin{abstract} 
The time evolution
of magnons  subject to a time-dependent microwave field is usually 
described within the so-called ``S-theory'', where
kinetic equations for the 
distribution function  are obtained within the time-dependent
Hartree-Fock approximation.  
To explain the recent observation of
``Bose-Einstein condensation of magnons'' in an external microwave field
[Demokritov {\it{et al.}}, Nature {\bf{443}}, 430 (2006)],
we extend the ``S-theory'' to include the Gross-Pitaevskii equation
for the time-dependent expectation values of the magnon creation and annihilation operators.
We explicitly solve the resulting coupled equations
within a simple  approximation where only a single condensed mode 
is retained.
We also  re-examine the usual derivation of an effective 
boson model from a realistic spin model
for  yttrium-iron garnet films
and argue that in the parallel pumping geometry (where
both the static and the  time-dependent magnetic field
are parallel to the macroscopic magnetization) 
the time-dependent Zeemann energy cannot give rise to 
magnon condensation.

\end{abstract}


\maketitle

\section{Introduction}
When ordered magnets are exposed to microwave radiation of sufficiently high power,
one typically observes 
an exponential growth of the population of  certain groups of spin-wave modes
during some intermediate time interval. This is an example for 
a  general phenomenon which is usually referred to as {\it{parametric resonance}}.
A particularly suitable system for observing parametric resonance are yttrium-iron garnet (YIG)
crystals, because the spin-waves in this system have a very low damping \cite{Cherepanov93}. 
Early microscopic theories explaining parametric resonance in magnetic insulators 
have  been developed by
Suhl\cite{Suhl57}, and by   Schl\"{o}mann and co-authors \cite{Schloemann60}.
In the 1970s  Zakharov, L'vov, and  Starobinets \cite{Zakharov70} developed 
a comprehensive kinetic theory of parametric resonance in magnon gases
which is sometimes called ``S-theory''. In this approach kinetic equations for 
the time-dependent distribution functions $n_{\bd{k}} ( t )  = \langle a^{\dagger}_{\bd{k}} ( t )
 a_{\bd{k}} ( t ) \rangle $ and
$p_{\bd{k}} ( t )  = \langle a_{-\bd{k}} ( t )
 a_{\bd{k}} ( t ) \rangle $ are derived within the self-consistent time-dependent Hartree-Fock approximation.
Here  $a_{\bd{k}} ( t )$ and  $a^{\dagger}_{\bd{k}} ( t )$ are the annihilation and creation operators of
magnons with momentum $\bd{k}$ in the Heisenberg picture.
Subsequently
the non-linear kinetic equations of the 
``S-theory'' and extensions thereof have been studied by
many authors \cite{Tsukernik75,Vinikovetskii79,Lim88,Kalafati89,Lvov94,Rezende09}. 

Quite recently  Demokritov and co-workers \cite{Demokritov06,Demidov07} observed a new
coherence effect of magnons  in YIG under the influence of an external microwave field
which they interpreted as Bose-Einstein condensation (BEC) of magnons at room temperature.
A similar phenomenon has been observed in superfluid $^3$He, where NMR pumping 
can cause the magnetization to precess phase-coherently \cite{Borovik84}.
The emergence of this coherent state can also be viewed as magnon BEC \cite{Volovik08,Bunkov09}.
Whether or not the experiments by Demokritov {\it{et al.}} \cite{Demokritov06,Demidov07} can be considered to be an analogue of BEC
in atomic Bose gases (which nowadays is routinely  realized using ultracold atoms in an optical trap)
has been discussed controversially in the literature \cite{Zvyagin07,Bugrij08}. We argue below that
the coherent state generated in these experiments \cite{Demokritov06,Demidov07} should perhaps 
not be called a Bose-Einstein condensate, because the condensation is not accompanied
by spontaneous symmetry breaking in this case; instead, the microwave field
gives rise to a term in the hamiltonian which
explicitly breaks the $U(1)$-symmetry of the magnon hamiltonian.

Unfortunately, the conventional ``S-theory'' is insufficient to describe the experimental situation,
because the coherent magnon state generated in the experiments is characterized by
finite expectation values of the
magnon annihilation and creation operators
 $a_{\bd{k}} ( t ) $
and  $ a^{\dagger}_{\bd{k}} ( t ) $ for certain special values of $\bd{k}$.
In the condensed phase, the kinetic equations for the pair correlators
$n_{\bd{k}} ( t )$ and $p_{\bd{k}} ( t )$ should therefore be augmented by
equations of motion for the expectation values
$ \langle a_{\bd{k}} ( t )  \rangle$
and  $ \langle a^{\dagger}_{\bd{k}} ( t )  \rangle$.
Recall that in the theory of the interacting Bose gas the corresponding
equation of motion for the order-parameter is called
Gross-Pitaevskii equation;~\cite{Pitaevskii03}
this equation is missing in the conventional ``S-theory'' which therefore 
does not completely describe the coherent magnon state in the regime of strong pumping.
In this work we shall outline an extension of  ``S-theory'' which includes the
order parameter dynamics on equal footing with the kinetic equations for the
distribution functions.
Since we would like to clarify conceptual points
rather than performing explicit quantitative calculations, we shall derive our
extended ``S-theory''
within the framework of a simple toy model which we motivate in the following section.

\section{Toy model for parametric resonance in YIG}

In order to understand a complex physical phenomenon, it is sometimes
useful to study a  simplified ``toy model'' which still contains some
essential features of  the phenomenon 
of interest.  For our purpose, it is sufficient to  consider a single
anharmonic oscillator with an additional time-dependent
term describing the creation and annihilation of pairs of particles.
The hamiltonian is 
\begin{eqnarray}
 \hat{H} ( t ) & = & 
\epsilon_0 a^{\dagger} a +  
\frac{\gamma_0}{2} e^{ -i \omega_0 t }  a^{\dagger} a^{\dagger} +
    \frac{\gamma_0^{\ast}}{2} e^{  i \omega_0 t }  a a  
 \nonumber
 \\ 
& + & \frac{u}{2} a^{\dagger} a^{\dagger} a a .
 \label{eq:Htoy}
 \end{eqnarray}
Here $a$ and $a^{\dagger}$ are bosonic annihilation and
creation operators, $\epsilon_0 > 0$ is 
some energy scale, and $u>0$ is the interaction energy.
The second and third terms on the right-hand side of Eq.~(\ref{eq:Htoy})
describe the effect of an external microwave-field which oscillates with frequency
$\omega_0 > 0$ and couples with strength $ \gamma_0  $ to the magnon gas. 
Below we shall show that this model contains the essential physics of 
parametric resonance and BEC of magnons; in particular,
in the regime of strong pumping $ |\gamma_0| > |\epsilon_0 - \omega_0/2 |$ 
the model has a stationary non-equilibrium state which corresponds  to  the
coherent magnon state observed in the experiments by Demokritov and 
co-workers \cite{Demokritov06,Demidov07}.

Our toy model (\ref{eq:Htoy}) 
involves only a single boson operator representing the magnon at the minimum
of the dispersion which is expected to condense.
Of course, for experimentally relevant macroscopic samples of YIG
a more realistic model should describe infinitely many magnon
operators $a_{\bd{k}}$ labeled by crystal-momentum $\bd{k}$, so that
the following bosonic ``resonance
hamiltonian'' should give
a better description of the experimental situation,
\begin{eqnarray}
 \hat{H}_{ \rm res } ( t ) & = & 
\sum_{\bd{k}} \epsilon_{\bd{k}} a_{\bd{k}}^{\dagger} a_{\bd{k}} 
 \nonumber
 \\
& + &  
\frac{1}{2} \sum_{\bd{k}}
\left[    
\gamma_{\bd{k}} e^{ -i \omega_0 t }  a^{\dagger}_{\bd{k}} a^{\dagger}_{ - \bd{k}} 
+ \gamma_{\bd{k}}^{\ast} e^{  i \omega_0 t }  a_{- \bd{k}} a_{\bd{k}}     
 \right]
 \nonumber
 \\ 
& + & \frac{1}{2} \sum_{ \bd{k}, \bd{k}^{\prime} , \bd{q}}
 u ( { \bd{k} ,   \bd{k}^{\prime}  , \bd{q} )
 a^{\dagger}_{\bd{k} + \bd{q}}}  a^{\dagger}_{ \bd{k}^{\prime} - \bd{q} }    a_{\bd{k}^{\prime}}
 a_{ \bd{k}} .
 \label{eq:Hpr}
 \end{eqnarray}
If we assume that the $\bd{k} =0$ boson condenses and retain only this degree of freedom
on the right-hand side of Eq.~(\ref{eq:Hpr}), we arrive at our toy model (\ref{eq:Htoy}). 
In the theory of superfluidity
a similar
reduced description involving only the order parameter
is provided by the Gross-Pitaevskii equation \cite{Pitaevskii03}.
Of course, the minimum  of the dispersion in experimentally relevant samples
of  YIG occurs at certain non-zero wave-vectors $ \pm \bd{k}_{\ast}$, so that
it would be more accurate to retain  the two modes $a_{ \bd{k}_{\ast} }$ and
$a_{  -\bd{k}_{\ast} }$ and their mutual interactions in Eq.~(\ref{eq:Hpr}).
Moreover, the fact that in the experiments \cite{Demokritov06,Demidov07}
the wave-vectors of the condensed magnons are
different from the wave-vectors of the magnons which are initially generated by
microwave pumping cannot be described within the framework of our toy model.
Nevertheless, below we shall show that
our simple model allows us to understand some conceptual
points related to the nature of the
coherent state observed in the experiments.~\cite{Demokritov06,Demidov07}

The bosonic resonance hamiltonian (\ref{eq:Hpr}) 
has been the starting point of several theoretical
investigations of parametric resonance in magnon gases 
\cite{Zakharov70,Tsukernik75,Vinikovetskii79,Lim88,Kalafati89,Lvov94,Rezende09}.
This model is believed to be a realistic model
for YIG in the parallel pumping geometry, 
where the static and the time-dependent components of the external magnetic fields 
are both parallel to the direction of the macroscopic magnetization.
In the appendix we shall critically re-examine
the usual derivation of Eq.~(\ref{eq:Hpr}) from an effective spin hamiltonian for YIG
and show that in spin language the time-dependent resonance term
in the second line of Eq.~(\ref{eq:Hpr}) involves also the combinations
$ \cos ( \omega_0 t )  [ S^x_i S^x_i - S^y_i S^y_i]$ and
$\sin ( \omega_0 t )  [ S^x_i S^y_i + S^y_i S^x_i ]$, where
$S^{\alpha}_i$ are the components of  the spin operators at lattice site $i$.
Terms of this type cannot be related to the  Zeemann energy
associated with a time-dependent magnetic field parallel to the magnetization.
This is  obvious for a ferromagnet with only exchange
interactions, because in this case the 
magnon operators
$a_{\bd{k}}$ and $a_{\bd{k}}^{\dagger}$ 
can be identified with the Fourier components of the
Holstein-Primakoff~\cite{Holstein40} bosons
 $a_i$ and $a^{\dagger}_i$, which in turn can be related to
the usual spin ladder operators $S^{+}_i$ and $S^{-}_i$; to leading order for 
large spin $S$,
 \begin{equation}
 S^{+}_i \approx \sqrt{2 S} a_i \; \; , \; \; S^{-}_i  \approx \sqrt{2 S} a^{\dagger}_i.
 \end{equation}
Note, however, that the spin Hilbert space  has only
$2S+1$ states per site, whereas the bosonic Fock space associated
with the canonical boson operators $a_i$ and $a^{\dagger}_i$
is  infinite dimensional;
the identification of magnons with canonical bosons is therefore only approximate.
For a description of
coherence phenomena involving  large occupancies of
magnon states one should therefore  keep in mind that
there is a constraint on the magnon Hilbert space.
Assuming for simplicity that the parameter
$\gamma_{\bd{k}} = \gamma$ in Eq.~(\ref{eq:Hpr})
is real and independent of $\bd{k}$, the second term on the right-hand side
of  Eq.~(\ref{eq:Hpr}) can be written as
\begin{eqnarray}
& &  \frac{ \gamma}{2} \sum_{\bd{k}}
\left[   
 e^{ -i \omega_0 t }  a^{\dagger}_{\bd{k}} a^{\dagger}_{ - \bd{k}} +
 e^{  i \omega_0 t }  a_{- \bd{k}} a_{\bd{k}}     
 \right]
 \nonumber
 \\
&    \approx  &
\frac{ \gamma }{4S} 
\sum_{i} \left[ 
e^{ - i \omega_0 t }  S_i^-  S_i^- +
e^{  i \omega_0 t } S_i^+ S_i^+    
\right]
\nonumber
 \\
&    =  &\frac{ \gamma }{2S} 
\sum_{i} \Bigl\{
 \cos ( \omega_0 t ) \left[ S_i^x  S_i^x - S_i^y S_i^y \right]
 \nonumber
 \\
  & & \hspace{10mm}
 - \sin ( \omega_0 t ) \left[ S_i^x  S_i^y + S_i^y  S_i^x \right] \Bigr\}.
 \label{eq:easyaxis}
\end{eqnarray}
In spin language, the pumping term in Eq.~(\ref{eq:Hpr}) therefore corresponds to 
a time-dependent single ion anisotropy whose easy axis  
rotates with frequency $\omega_0$ around the $z$-axis. 
Of course,  the magnon operators for
YIG are not directly related to Holstein-Primakoff bosons because
an additional Bogoliubov transformation is necessary to diagonalize  
the quadratic part of the boson hamiltionian. Nevertheless, we show in the
appendix that also in this case the pumping term
in the effective boson hamiltonian (\ref{eq:Hpr}) can be related to a rotating
easy axis anisotropy of the above type.

\section{Kinetic equations}
To discuss the time evolution  of  our toy model defined in Eq.~(\ref{eq:Htoy})
it is convenient to remove the explicit time dependence from the
hamiltonian  $\hat{H} ( t )$ by performing a canonical transformation to the
``rotating reference frame'', 
 \begin{subequations}
 \begin{eqnarray}
 \tilde{a} & = &  e^{   \frac{i}{2}  \omega_0 t} a  =
 \hat{U}_0 ( t ) a \hat{U}_0^{\dagger} ( t ) ,
 \\
 \tilde{a}^{\dagger} & = &  e^{  -  \frac{i}{2}  \omega_0 t  } a^{\dagger} =
\hat{U}_0 ( t ) a^{\dagger} \hat{U}^{\dagger}_0 ( t )  ,
 \end{eqnarray}
 \end{subequations}
where $\hat{U}_0 ( t ) = e^{ - \frac{i}{2}  \omega_0 t a^{\dagger} a }$.
The new operators satisfy the Heisenberg equations of motion
 \begin{equation}
 i \partial_t \tilde{a} = [ \tilde{a} , \tilde{H} ] \; \; ,
 \; \;  i \partial_t \tilde{a}^{\dagger} = [ \tilde{a}^{\dagger} , \tilde{H} ],
 \label{eq:HEOM} 
\end{equation}
where the rotated hamiltonian $\tilde{H}$ of our toy model
does not depend explicitly on time,
 \begin{eqnarray}
 \tilde{H}  & = & 
\tilde{\epsilon}_0 \tilde{a}^{\dagger} \tilde{a} +  
\frac{\gamma_0}{2}   \tilde{a}^{\dagger} \tilde{a}^{\dagger} 
  +  \frac{\gamma_0^{\ast}}{2}   \tilde{a} \tilde{a}  +    
\frac{u}{2} \tilde{a}^{\dagger} \tilde{a}^{\dagger} \tilde{a} \tilde{a} .
 \label{eq:Htoyrot}
 \end{eqnarray}
Here we have introduced the shifted oscillator energy
 \begin{equation}
  \tilde{\epsilon}_0 = \epsilon_0 - \frac{\omega_0}{2}.
 \end{equation}
To relate correlation functions in the original model to those in the
rotating frame, we simply have to insert the appropriate phase factors.
For example, in ``S-theory'' one usually considers the normal distribution function,
 \begin{equation}
 n ( t ) =  \langle a^{\dagger} ( t ) a (t) \rangle = \langle \tilde{a}^{\dagger} ( t ) \tilde{a} (t) \rangle,
 \end{equation}
and its anomalous counter-part,
\begin{equation}
 p ( t ) =  \langle a ( t ) a (t) \rangle = e^{ - i \omega_0 t}
 \langle \tilde{a} ( t ) \tilde{a} (t) \rangle   \equiv   e^{ - i \omega_0 t}  \tilde{p} ( t) ,
 \end{equation}
where
 expectation values are with respect to some density matrix
$\hat{\rho} ( t_0 ) $ specified at time $t_0$,
 \begin{equation}
 \langle \ldots \rangle = {\rm Tr} [ \hat{\rho} ( t_0 ) \ldots ].
 \end{equation}
Throughout this work we shall mark all quantities defined in the rotating reference frame 
by a tilde.

\subsection{Instability of the non-interacting system}

In the non-interacting limit $(u=0)$ the equations of motion 
for the distribution functions $n ( t )$ and $\tilde{p} ( t )$ can be obtained
trivially from the equations of motion (\ref{eq:HEOM}) of the operators
$\tilde{a} (t)$ and $\tilde{a}^{\dagger} ( t )$ in the rotating reference frame,
\begin{subequations}
 \begin{eqnarray}
 i \partial_t n ( t ) & = &   {\gamma}_0  \tilde{p}^{\ast} (t ) -  \gamma_0^{\ast} 
\tilde{p} (t) ,
 \label{eq:kinn1}
 \\
 i  \partial_t \tilde{p} ( t ) & = & 2 \tilde{\epsilon}_0 \tilde{p} (t )  +
   {\gamma}_0  [ 2 n (t ) + 1 ] .
 \label{eq:kinp1}
\end{eqnarray}
 \end{subequations}
These equations can be solved exactly. 
For $|\tilde{\epsilon}_0 | > | {\gamma}_0 |$ the solution is oscillatory,
while in the strong pumping regime
$|{\gamma}_0 | > | \tilde{\epsilon}_0 |$ the solutions grow exponentially.
Let us explicitly give the 
solution of Eqs.~(\ref{eq:kinn1},\ref{eq:kinp1}) with initial conditions
$ n(0 ) = n_0$ and $\tilde{p} ( 0 ) =0$. For simplicity, we assume in the rest of this
work that $\gamma_0$ is real and positive;
the case of complex 
$\gamma_0 = | \gamma_0 | e^{i \varphi}$ can be reduced to real $\gamma_0 > 0$
by absorbing the phase factor $e^{i \varphi}$
into a re-definition of the anomalous correlator, 
$e^{ - i \varphi} \tilde{p} ( t ) \rightarrow \tilde{p} ( t )$. 
Defining
 \begin{equation}
  \alpha \equiv   \sqrt{  \tilde{\epsilon}_0^2  -   \gamma_0^2 } ,
 \label{eq:alphadef} 
\end{equation}
the solution in the weak pumping regime  $ \gamma_0 < | \tilde{\epsilon}_0 |$
can be written as
 \begin{subequations}
\begin{eqnarray}
 \frac{ {\rm Re} \tilde{p} ( t )}{ n_0 + \frac{1}{2}} & = & 
-    \gamma_0  \tilde{\epsilon}_0  \frac{ 1 - \cos ( 2 \alpha t )}{
 \alpha^2 }, 
 \label{eq:freeosc1}
\\
\frac{{\rm Im} \tilde{p} (t ) }{ n_0 + \frac{1}{2}} & = & -     \gamma_0  \frac{\sin ( 2 \alpha t )}{  \alpha  },
 \label{eq:freeosc2}
\\
\frac{n ( t ) + \frac{1}{2} }{n_0 + \frac{1}{2}} & = & 1 +     \gamma_0^2 
\frac{ 1 - \cos ( 2 \alpha t )}{
 \alpha^2 }.
 \label{eq:freeosc3}
 \end{eqnarray}
\end{subequations}
In the opposite strong pumping regime 
$  \gamma_0  >  | \tilde{\epsilon}_0 |$ the solution
can be obtained by replacing $\alpha \rightarrow i \beta$ in the above expressions,
where
 \begin{equation}
 \beta =  \sqrt{  \gamma_0^2 -    \tilde{\epsilon}_0^2 }. 
\label{eq:betadef} 
\end{equation}
Then we obtain
 \begin{subequations}
 \begin{eqnarray}
 \frac{ {\rm Re} \tilde{p} ( t )}{ n_0 + \frac{1}{2}} & = & 
-    \gamma_0  \tilde{\epsilon}_0  \frac{ \cosh ( 2 \beta t )-1}{
 \beta^2 }, 
\label{eq:freeres1}
\\
\frac{{\rm Im} \tilde{p} (t ) }{ n_0 + \frac{1}{2}} & = &
 -     \gamma_0  \frac{\sinh ( 2 \beta t )}{  \beta  },
\label{eq:freeres2}
 \\
\frac{n ( t ) + \frac{1}{2} }{n_0 + \frac{1}{2}} & = &
1 +     \gamma_0^2 
\frac{  \cosh ( 2 \beta t ) - 1}{
 \beta^2 }.
\label{eq:freeres3}
 \end{eqnarray}
\end{subequations}
The behavior at the threshold value $ \gamma_0 = | \tilde{\epsilon}_0 |$
can be obtained either from Eqs.~(\ref{eq:freeosc1}--\ref{eq:freeosc3}) 
for $\alpha \rightarrow 0$, or
from
Eq.~(\ref{eq:freeres1}--\ref{eq:freeres3}) for $\beta \rightarrow 0$,
\begin{subequations}
 \begin{eqnarray}
 \frac{ {\rm Re} \tilde{p} ( t )}{ n_0 + \frac{1}{2}} & = & 
-    2 \gamma_0  \tilde{\epsilon}_0  t^2,
\label{eq:thres1}
\\
\frac{{\rm Im} \tilde{p} (t ) }{ n_0 + \frac{1}{2}} & = &
 -     2 \gamma_0  t,
\label{eq:thres2}
 \\
\frac{n ( t ) + \frac{1}{2} }{n_0 + \frac{1}{2}} & = &
1 +     2 \gamma_0^2 t^2.
\label{eq:thres3}
 \end{eqnarray}
\end{subequations}
Physically, the  exponential increase of correlations 
for $\gamma_0 > |\tilde{\epsilon}_0|$
is a consequence of the fact that in this regime the non-interacting part of the
hamiltonian $\tilde{H}$  in Eq.~(\ref{eq:Htoyrot}) is not bounded from below.
This is easily seen by setting 
 \begin{equation}
\tilde{a} = \frac{ \hat{X} + i \hat{P} }{\sqrt{2}}
 \; \; , \; \; 
 \tilde{a}^{\dagger} = \frac{ \hat{X} - i \hat{P} }{\sqrt{2}},
 \end{equation}
so that
 \begin{equation}
\tilde{\epsilon}_0 \tilde{a}^{\dagger} \tilde{a} +  
    \frac{\gamma_0}{2}   [
  \tilde{a}^{\dagger} \tilde{a}^{\dagger}  + \tilde{a} \tilde{a}      ] = 
 \frac{\tilde{\epsilon}_0 - \gamma_0}{2}   \hat{P}^2
+ \frac{ \tilde{\epsilon}_0 + \gamma_0}{2}  \hat{X}^2.
 \end{equation}
Obviously, for $\gamma_0  > | \tilde{\epsilon}_0 |$ the non-interacting
part of our toy model describes a harmonic oscillator with negative mass.
The spectrum of such a quantum mechanical system is not bounded from below,
which gives rise to the exponential growth of correlations discussed above.
Fortunately, this pathology of the non-interacting limit is cured for any positive value of
the interaction.  The physical consequences of this are most transparent if we consider
the equations of motion for the expectation values of the creation and annihilation operators,
which will be discussed in the  following subsection.

\subsection{Gross-Pitaevskii equation}

\begin{figure}[tb]
  \centering
\includegraphics[width=75mm]{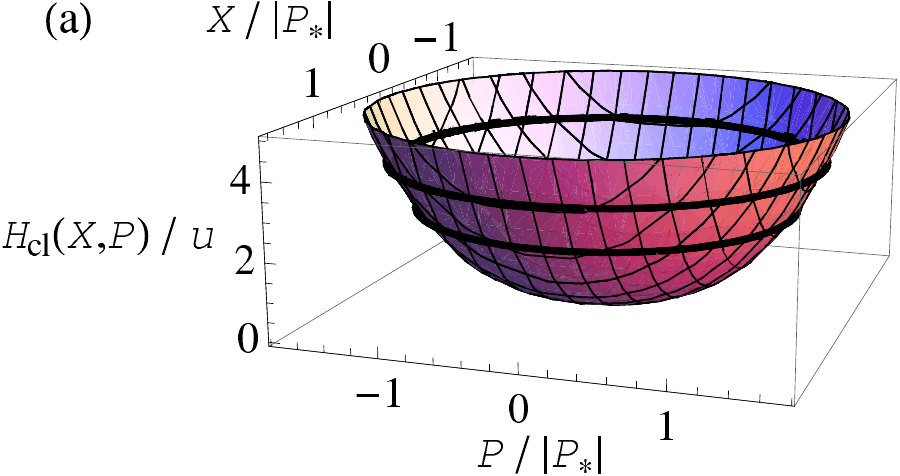}
\includegraphics[width=75mm]{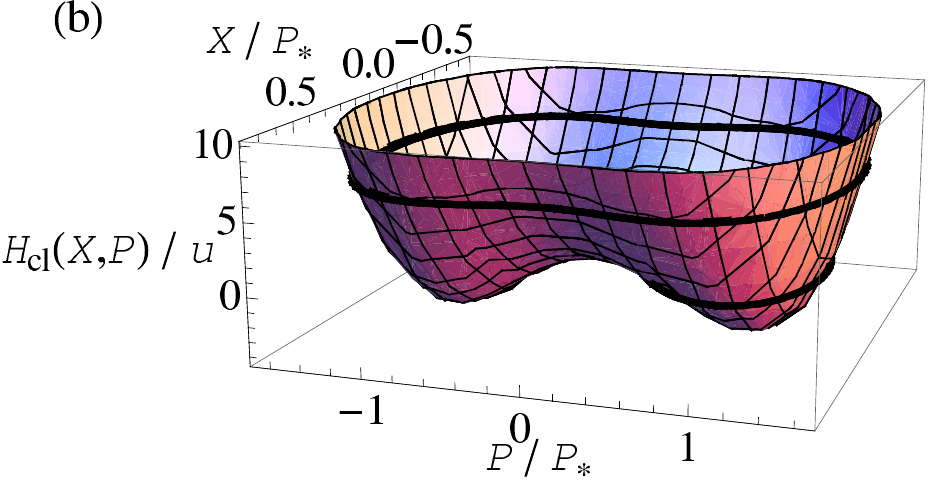}
  \vspace{-4mm}
  \caption{%
(Color online)
Graph of the
classical hamiltonian $H_{\rm cl} ( X , P )$ 
defined in Eq.~(\ref{eq:Hcl}). The corresponding classical hamiltonian
equations of motion are equivalent to the
Gross-Pitaevskii equation (\ref{eq:GP1}) for the
complex order parameter $\phi ( t ) = ( X ( t ) + i P(t))/\sqrt{2}$.
The thick black lines are solutions of the equations of motion  for different initial conditions.
$X$ and $P$ are both measured in units of the momentum scale $ |P_{\ast} | = 
\sqrt{ 2   |\gamma_0 - \tilde{\epsilon}_0 | / u }$.
(a):~$\tilde{\epsilon}_0/u = 10$ and
$\gamma_0 / u =2$; note that for
$|\tilde{\epsilon}_0| >  \gamma_0 $ our classical hamiltonian $H_{\rm cl} ( X , P )$ 
has a global minimum for $X=P =0$. 
(b):~$\tilde{\epsilon}_0/u = 10$ and
$\gamma_0 / u =40$;
in the regime
  $  \gamma_0  >   |\tilde{\epsilon}_0|$
our classical hamiltonian has two degenerate minima at $(X , P ) = (0, \pm P_{\ast})$, 
so that the graph  of $H_{\rm cl} ( X , P )$ 
shown in (b)
has some similarity to the shape of Napoleon's hat\cite{Amore09}. 
}
  \label{fig:potential}
\end{figure}

The toy model hamiltonian (\ref{eq:Htoyrot}) in the rotating reference frame 
gives rise to
the following Heisenberg equation of motion 
for the annihilation operator,
\begin{equation}
 i \partial_t  \tilde{a}   =  \tilde{\epsilon}_0   \tilde{a} 
 + \gamma_0  \tilde{a}^{\dagger} + u   \tilde{a}^{\dagger} \tilde{a}^2  .
 \end{equation}
Taking the expectation value of both sides and
factorizing the expectation value of the interaction term as follows,
 \begin{equation}
 \langle \tilde{a}^{\dagger} \tilde{a}^2 \rangle \rightarrow
\langle \tilde{a}^{\dagger} \rangle \langle \tilde{a} \rangle^2 ,
 \label{eq:replace}
 \end{equation}
we obtain the Gross-Pitaevskii equation for the time-dependent
order-parameter
$\phi ( t ) \equiv \langle \tilde{a} (t) \rangle$
in the rotating reference frame,
\begin{eqnarray}
 i \partial_t \phi & = & \tilde{\epsilon}_0  \phi
 + \gamma_0 \phi^{\ast} + u   | \phi |^2 \phi
= \frac{ \partial H_{\rm cl} ( {\phi}^{\ast} , \phi )}{ \partial {\phi}^{\ast} },
 \label{eq:GP1}
\end{eqnarray}
where the effective classical hamiltonian 
$H_{\rm cl}$ is given by
\begin{equation}
  H_{\rm cl} ( {\phi}^{\ast} , \phi ) = \tilde{\epsilon}_0 |  \phi |^2 + 
 \frac{\gamma_0}{2}  [ \phi^{\ast 2} +    {\phi}^2 ] + 
\frac{u}{2} | \phi |^4.
 \end{equation} 
Writing $\phi =  ( X + i P )/\sqrt{2}$ we may alternatively write
\begin{equation}
  H_{\rm cl} ( X , P )
 = \frac{ \tilde{\epsilon}_0 -  \gamma_0 }{2} P^2 +
 \frac{ \tilde{\epsilon}_0 +  \gamma_0 }{2} X^2 +
 \frac{u}{8} ( X^2 + P^2 )^2.
 \label{eq:Hcl}
 \end{equation}
Because the classical hamiltonian $H_{\rm cl}  ( X ( t ), P ( t ))$  is conserved
along the flow defined by the
Gross-Pitaevskii equation, 
the solutions of Eq.~(\ref{eq:GP1}) are simply given by the curves 
of constant
$H_{\rm cl} ( X ( t ) , P ( t ))$ in phase space.
The shape of $H_{\rm cl}$ and typical trajectories
are shown in Fig.~\ref{fig:potential}.
Note that in the strong pumping regime $  \gamma_0  >   |\tilde{\epsilon}_0|$
the function
 $H_{\rm cl} ( X , P )$ has two degenerate minima at
 \begin{equation}
 X = 0 \; \; , \; \; P = \pm P_{\ast} = \pm
\sqrt{ \frac{2  (\gamma_0 - \tilde{\epsilon}_0 ) }{ u }}, 
\end{equation}
corresponding to stationary points (in the rotating reference frame) of the system. 
Note that at these special points the
expectation value of the annihilation  operator is purely imaginary,
 \begin{equation}
 \langle \tilde{a} \rangle = \pm \frac{i}{\sqrt{2}}  P_{\ast}
=
\pm i \sqrt{ \frac{  \gamma_0 - \tilde{\epsilon}_0 }{ u } } .
 \end{equation}
The associated stationary
points of the dynamical system (\ref{eq:GP1})
describe a coherent magnon state where the macroscopic magnetization
has a rotating component perpendicular to the static magnetic field.
In bosonic language, such a state corresponds to a coherent state,
which is an eigenstate of the annihilation operator~\cite{Rezende09,Araujo74}.
Whether or not this state should be called a Bose-Einstein condensate
of magnons seems to be a semantic question.
In our opinion  this  terminology is somewhat misleading, because
this coherent magnon state does not exhibit 
spontaneous symmetry breaking which is one of  the
most important properties of a Bose-Einstein condensate in 
interacting Bose gases. Instead, the coherent magnon state 
observed by Demokritov and co-workers\cite{Demokritov06,Demidov07}
is generated by an external pumping field which explicitly breaks
the $U ( 1 )$-symmetry of the magnon hamiltonian.
In the static limit, the role of a similar symmetry breaking term on the
Bose-Einstein condensation of magnons has recently been discussed
by Dell'Amore, Schilling, and Kr\"{a}mer \cite{Amore09}.

\subsection{Time-dependent Hartree-Fock approximation}

\begin{figure}[tb]
  \centering
\includegraphics[width=70mm]{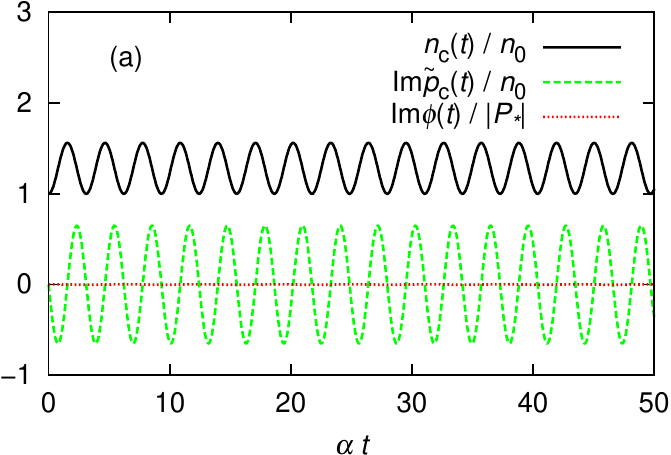}
\includegraphics[width=70mm]{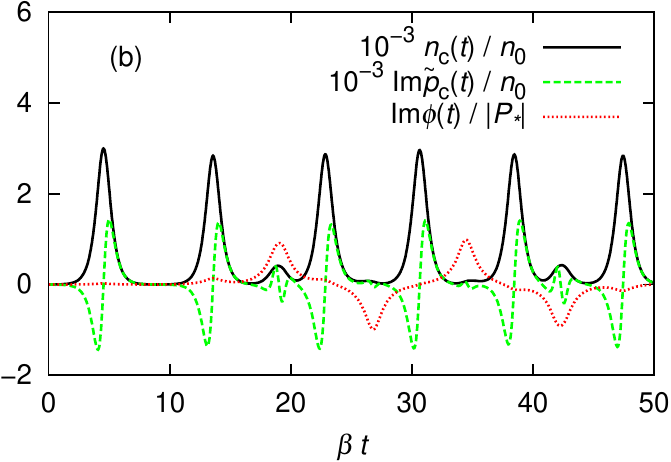}
\includegraphics[width=70mm]{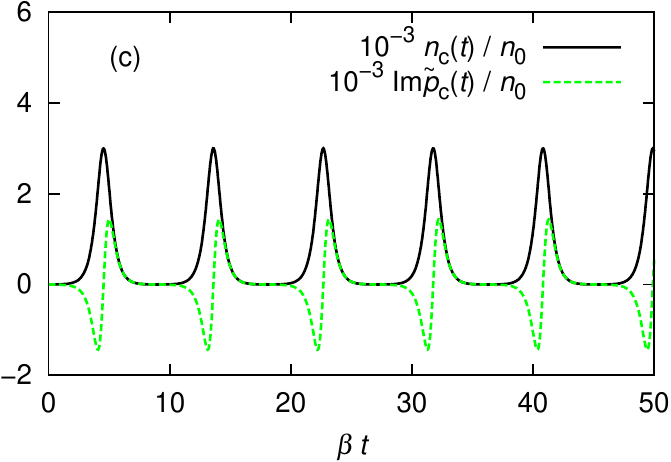}
  \vspace{-4mm}
  \caption{%
(Color online) Numerical solution of the coupled kinetic 
equations (\ref{eq:GP2}, \ref{eq:kinn2}, \ref{eq:kinp2})
with initial conditions $n_c ( 0 ) = n_0 = 1$,
 $\tilde{p}_c (0) =0$, and $\phi ( 0 )  = 0.05 i$.
The characteristic energy scales $\alpha$ and $\beta$ are defined in
Eqs. (\ref{eq:alphadef}, \ref{eq:betadef}).
(a):~$\tilde{\epsilon}_0/u =500$ and $ \gamma_0/u =200$.
Recall that in the absence of interactions there is no instability as long as 
$ |\tilde{\epsilon}_0 | >  \gamma_0 $. 
(b):~$\tilde{\epsilon}_0/u =500$ and $ {\gamma}_0/u =5000$.
In this regime there would be  an instability in the non-interacting limit,
but in the interacting system all correlations remain finite.
(c):~Same parameters as (b) but without finite expectation values, like in
the conventional ``S-theory''.
}
  \label{fig:kinsolution}
\end{figure}

Let us now take into account
the leading fluctuation correction to the replacement
(\ref{eq:replace})  in the derivation of the
Gross-Pitaevskii equation (\ref{eq:GP1}).
To first order in $u$, fluctuations simply renormalize the bare parameters
$\tilde{\epsilon}_0$ and $\gamma_0$ in Eq.~(\ref{eq:GP1}) as follows,
 \begin{subequations} 
 \begin{eqnarray}
 \tilde{\epsilon}_0 & \rightarrow & \tilde{\epsilon}_c ( t )  =  \tilde{\epsilon}_0 + 2 u n_c ( t ) ,
 \label{eq:epst2}
 \\
  \gamma_0 & \rightarrow & {\gamma}_c ( t ) =  \gamma_0  + u \tilde{p}_c ( t ),
 \label{eq:gammat2}
 \end{eqnarray}
 \end{subequations}
where the connected correlation functions 
$n_c (t )$ and $\tilde{p}_c (  t )$ in the rotating reference frame are defined by
 \begin{subequations} 
\begin{eqnarray}
 n_c ( t ) & = & \langle \delta \tilde{a}^{\dagger} ( t ) \delta \tilde{a} ( t ) \rangle,
 \\
\tilde{p}_c ( t ) & = & \langle \delta \tilde{a} ( t ) \delta \tilde{a} ( t ) \rangle,
 \end{eqnarray}
\end{subequations}
with $\delta \tilde{a} (t ) = \tilde{a} ( t ) - \langle \tilde{a} (t ) \rangle$.
Instead of the Gross-Pitaevskii equation (\ref{eq:GP1}) we now obtain for the
order parameter dynamics,
\begin{equation}
 i \partial_t \phi  =  \tilde{\epsilon}_c ( t )  \phi
 + \gamma_c (t ) \phi^{\ast} + u   | \phi |^2 \phi .
 \label{eq:GP2}
 \end{equation}
Note that this generalized Gross-Pitaevskii equation depends on the
connected correlation functions $n_c( t )$ and $\tilde{p}_c( t )$, which we
calculate in self-consistent Hartree-Fock approximation.
The resulting equations of motion can be obtained from the
corresponding non-interacting kinetic equations (\ref{eq:kinn1},\ref{eq:kinp1})
by substituting
 \begin{subequations}
\begin{eqnarray}
 \tilde{\epsilon}_0  \rightarrow \tilde{\epsilon} ( t )  & = &  \tilde{\epsilon}_0 
+ 2 u [    n_c ( t ) + | \phi ( t ) |^2 ],
 \label{eq:epst3}
 \\
  \gamma_0  \rightarrow  {\gamma} ( t ) & = & \gamma_0  + u [
\tilde{p}_c ( t ) +\phi^2 ( t )  ].
 \label{eq:gammat3}
 \end{eqnarray}
 \end{subequations}
The kinetic equations for the connected distribution functions are therefore
\begin{subequations}
 \begin{eqnarray}
 i \partial_t n_c ( t ) & = &   {\gamma} ( t )  \tilde{p}_c^{\ast} (t ) -  \gamma^{\ast} ( t ) 
\tilde{p}_c (t) ,
 \label{eq:kinn2}
 \\
 i  \partial_t \tilde{p}_c ( t ) & = & 2 \tilde{\epsilon} ( t ) \tilde{p}_c (t )  +
   {\gamma} (t )  [ 2 n_c (t ) + 1 ] .
 \label{eq:kinp2}
\end{eqnarray}
 \end{subequations}
For $\phi=0$ these equations reduce to the kinetic equations
obtained within ``S-theory'' \cite{Zakharov70}.
The numerical solution of Eqs.~(\ref{eq:GP2}, \ref{eq:kinn2}, \ref{eq:kinp2})
for $n_c ( 0 ) = n_0$, $\tilde{p}_c (0) =0$, and  infinitesimal ${\rm Im} \phi ( t ) > 0$ 
is shown in Fig.~\ref{fig:kinsolution}.
Obviously,  for sufficiently strong pumping an infinitesimal initial value of
$\phi ( 0 )$ builds up to a finite oscillation. Moreover, 
the connected correlation functions $n_c ( t )$ and $\tilde{p}_c ( t )$
remain always bounded, in contrast to the
exponentially growing correlations 
in the non-interacting limit given in Eqs.~(\ref{eq:freeres1}--\ref{eq:freeres3}).
Note also that the time evolution of 
the connected correlation functions 
appears to be rather irregular as soon as
the order-parameter has built up to a finite value.
In the conventional ``S-theory'' the quantities $n_c$ and $\tilde{p}_c$
are periodic (Fig.~\ref{fig:kinsolution}c), while including
the order parameter dynamics disturbes this strict periodicity (Fig.~\ref{fig:kinsolution}b).
This feature is still missing within the usual ``S-theory''in the strong pumping regime.

\section{Summary and conclusions}
Let us briefly summarize the two main results of this work:

First of all, we have shown that a complete theoretical description
of the coherent magnon state emerging in YIG for
sufficiently strong microwave pumping
requires an extension of the usual ``S-theory'' which includes the
Gross-Pitaevskii type of equation for the expectation values of the magnon operators.
Within a simple toy model consisting only of a single magnon mode we
have shown how to construct such an extension.
The explicit
solution of the resulting kinetic equations
shows that the order parameter dynamics
strongly influences the distribution functions.
 
Our second main result is the observation that
in spin-language 
the usual bosonic resonance 
hamiltonian (\ref{eq:Hpr}) corresponds to  a 
time-dependent rotating easy axis anisotropy
whose axis is perpendicular to the direction of the external field.
If this anisotropy is sufficiently strong, it gives rise to a 
forced oscillation of the macroscopic magnetization  around the
direction of the static external field. 
Although this phenomenon can be described in terms of  a coherent magnon state,
it should not be called a Bose-Einstein condensate, because
the emergence of this state is not associated with any kind of  spontaneous symmetry
breaking.

In future work, we shall further extend our approach in two directions:
on the one hand, a realistic model for YIG involves a quasi-continuum of magnon modes,
which can condense at finite wave-vectors $\pm {\bd{k}}_{\ast}$. 
For a more realistic quantitative description of the experiments,
we should therefore generalize our extended ``S-theory'' to include
all magnon modes relevant to the experiments on YIG.
This would also allow us to distinguish between the
``primary magnons'' created by the external pumping, and the
``condensing magnons'' with wave-vectors 
at the minima of the dispersion.
The second direction for improving our approach is to include
correlation effects beyond
the self-consistent Hartree-Fock approximation
into the kinetic equations.
For example,
to second order in $u$ the kinetic equations will contain relaxation terms
which will damp the oscillatory time dependence
found at the Hartree-Fock level.
Work in both directions is in progress.

\section*{ACKNOWLEDGMENTS}
We thank L. Bartosch and A. Isidori for discussions and
acknowledge financial support by
SFB/TRR49 and the DAAD/CAPES PROBRAL-program.
The work by A.~K. and P.~K. was partially carried out at the
International Center for Condensed Matter Physics at the University
of Bras\'{\i}lia, Brazil. We thank A. Ferraz for his 
hospitality. We are grateful to Y.~M. Bunkov and G.~E. Volovik for drawing our 
attention to analogies between experiments on YIG and superfluid $^3$He, and to
A. A. Zvyagin for sending us a copy of Ref.~[\onlinecite{Zvyagin82}].

\begin{appendix}
\renewcommand{\theequation}{A\arabic{equation}}
\section*{APPENDIX: PARALLEL PUMPING OF MAGNONS IN YIG}
\setcounter{equation}{0}

It is generally accepted that 
 the magnetic properties of YIG in the  parallel pumping geometry  
can be modelled by the following time-dependent quantum 
spin model\cite{Tupitsyn08,Kreisel09},
\begin{eqnarray}
  \hat{H}_{\rm YIG} (t )  & = & -\frac 12 \sum_{ij}\sum_{\alpha\beta} \left[J_{ij}\delta^{\alpha\beta}+D_{ij}^{\alpha\beta}\right] S_i^\alpha S_j^\beta 
 \nonumber
 \\
 & & 
- [ h_0 + h_1 \cos ( \omega_0 t ) ] 
\sum_i S_i^z,
 \label{eq:HYIG}
\end{eqnarray}
where $\alpha, \beta = x,y,z$ label the three spin components, and
the exchange couplings $J_{ij} = J( \bd{r}_i - \bd{r}_j )$
are only finite if the lattice sites $\bd{r}_i$ and $\bd{r}_j$ 
are nearest neighbors on a cubic lattice
with lattice spacing $a \approx 12.376 \text{\AA}$.
The value of the nearest neighbor exchange is $J \approx 1.29$K.
The dipolar tensor
$D_{ij}^{\alpha \beta}=D^{\alpha\beta}(\bd r_i-\bd r_j)$ is explicitly
\begin{eqnarray}
 D_{ij}^{\alpha\beta}&=&(1-\delta_{ij})\frac{\mu^2}{|\bd r_{ij}|^3}\left[3\hat{r}_{ij}^\alpha\hat{r}_{ij}^\beta-\delta^{\alpha\beta}\right],
\label{eq:defdip}
\end{eqnarray}
where 
$\bd{r}_{ij} = \bd{r}_i - \bd{r}_j$ and
$\hat{\bd{r}}_{ ij} = \bd{r}_{ ij } / | \bd{r}_{ij} |$.
 If we arbitrarily set the magnetic moment $\mu = 2 \mu_B = e \hbar /(mc)$, 
then we should work with an effective spin $S \approx 14.2$, as discussed
in Ref.~[\onlinecite{Kreisel09}].
Here $h_0 $ and $h_1$ are the amplitudes of the static and oscillating magnetic field
(multiplied by $\mu$). We assume that $h_0 > | h_1 |$ and that
both the static and the oscillating
magnetic field point into the 
direction of the macroscopic magnetization which we call the $z$-axis.
At this point one might already wonder how in this parallel pumping
geometry one can possibly arrive at a bosonic resonance hamiltonian of the form
(\ref{eq:Hpr}), which according to Eq.~(\ref{eq:easyaxis}) can be related
to some rotating easy axis anisotropy.  In fact, we shall show shortly that
the spin hamiltonian (\ref{eq:HYIG}) with parallel pumping {\it{cannot}} 
be reduced to the bosonic resonance hamiltonian (\ref{eq:Hpr}).

To bosonize the hamiltonian (\ref{eq:HYIG}) we express the spin operators
in terms of boson operators $b_i$ and $b_i^{\dagger}$ by means 
Holstein-Primakoff transformation,\cite{Holstein40}
 \begin{subequations} 
\begin{eqnarray}
 S_i^{+} & =  & \sqrt{2S} \sqrt{ 1 - \frac{ b_i^{\dagger} b_i}{2S}} b_i  = ( S_i^- )^{\dagger}   ,
 \\ 
S_i^z & = & S - b^{\dagger}_i b_i.
 \end{eqnarray}
 \end{subequations}
As usual, the  square roots are then expanded in powers of $1/S$, resulting
in a hamiltonian of the form
 \begin{equation}
  \hat{H}_{\rm YIG} (t ) = H_0 ( t ) + \hat{H}_2  ( t ) + \hat{H}_{ \rm int},
 \end{equation}
where $H_0 ( t )$ is a time-dependent constant,
$\hat{H}_2  ( t )$ is quadratic in the boson operators, 
 and the time-independent interaction
$\hat{H}_{ \rm int}$ involves three and more boson operators.
After Fourier transformation to momentum space the quadratic part of the
hamiltonian can be written as
 \begin{eqnarray}
\hat{H}_2 ( t ) & = & \sum_{\bd{k}} 
 \left[ A_{\bd{k}} b^{\dagger}_{\bd{k}} b_{\bd{k}} +
\frac{B_{\bd{k}}}{2} b^{\dagger}_{ \bd{k}} b^{\dagger}_{-\bd{k}} +
 \frac{B^{\ast}_{\bd{k}}}{2} b_{- \bd{k}} b_{\bd{k}}   
 \right]
 \nonumber
 \\
 & & + h_1 \cos ( \omega_0 t ) \sum_{\bd{k}} b^{\dagger}_{\bd{k}} b_{\bd{k}}.
 \end{eqnarray}
where
 \begin{subequations}
 \begin{eqnarray}
 A_{\bd{k}} & = & A_{ - \bd{k}}  = \sum_i e^{ - i \bd{k} \cdot \bd{r}_{ ij} } A_{ij},
 \\
B_{\bd{k}} & = & B_{ - \bd{k}}  = \sum_i e^{ - i \bd{k} \cdot \bd{r}_{ ij} } B_{ij},
\end{eqnarray}
\end{subequations}
with
 \begin{subequations}
 \begin{eqnarray}
 A_{ij} & = & \delta_{ij} h_0 + S ( \delta_{ij} \sum_{n} J_{ in} - J_{ ij} )
 \nonumber
 \\ 
 & + & S \left[ \delta_{ij} \sum_{n} D_{in}^{zz}-
\frac{D_{ij}^{xx} + D_{ij}^{yy}}{2} \right],
 \\
 B_{ij} & = & - \frac{S}{2}  \left[ D_{ij}^{xx} + 2 i  D_{ij}^{xy} - D_{ij}^{yy} \right].
 \end{eqnarray}
 \end{subequations}
Finally, we use a Bogoliubov transformation to diagonalize the
time-independent part of $\hat{H}_2 ( t )$,
\begin{eqnarray}
 \left( \begin{array}{c}
  b_{\bd{k}} \\
 b^{\dagger}_{-\bd{k}}
 \end{array}
 \right) = \left( \begin{array}{cc} u_{\bd{k}} & - v_{\bd{k}} \\
 -  v_{\bd{k}}^{\ast} & u_{\bd{k}}  \end{array} \right) 
\left( \begin{array}{c}
  a_{\bd{k}} \\
 a^{\dagger}_{-\bd{k}}
 \end{array}
 \right),
 \label{eq:Bogoliubov}
\end{eqnarray}
where
 \begin{equation}
 u_{\bd{k}} = \sqrt{ \frac{ A_{\bd{k}} + \epsilon_{\bd{k}} }{ 2 \epsilon_{\bd{k}} } } 
 \; \;   , \; \;
 v_{\bd{k}} = \frac{ B_{\bd{k}}}{ | B_{\bd{k}}    | }
\sqrt{ \frac{ A_{\bd{k}} - \epsilon_{\bd{k}} }{ 2 \epsilon_{\bd{k}} } } ,
 \end{equation}
and
 \begin{equation}
 \epsilon_{\bd{k}} = \sqrt{ A_{\bd{k}}^2 - | B_{\bd{k}} |^2 }.
 \end{equation}
After this transformation the hamiltonian reads\cite{Lvov94}
 \begin{eqnarray}
\hat{H}_2 ( t ) &= & \sum_{\bd{k}} 
 \left[ \epsilon_{\bd{k}} a^{\dagger}_{\bd{k}} a_{\bd{k}}  + \frac{ \epsilon_{\bd{k}}
 - A_{\bd{k}} }{2} \right]
 \hspace{30mm}
 \nonumber
 \\
 &  & \hspace{-15mm} + h_1 \cos ( \omega_0 t )  \sum_{\bd{k}} 
 \left[ \frac{A_{\bd{k}}}{\epsilon_{\bd{k}}} a^{\dagger}_{\bd{k}} a_{\bd{k}} +   \frac{A_{\bd{k}} - \epsilon_{\bd{k}} }{
2 \epsilon_{\bd{k}} } \right]
\nonumber
 \\
& & \hspace{-15mm} + \sum_{\bd{k}} \left[ 
 \gamma_{\bd{k}}
\cos ( \omega_0 t ) 
  a^{\dagger}_{\bd{k}} a^{\dagger}_{-\bd{k}} 
+
\gamma_{\bd{k}}^{\ast}
\cos ( \omega_0 t ) 
a_{-\bd{k}} a_{\bd{k}}
 \right],
 \label{eq:Hparpump}
 \end{eqnarray}
where
 \begin{equation}
 \gamma_{\bd{k}} = - \frac{h_1 B_{\bd{k}}}{ 2 \epsilon_{\bd{k}}} .
 \end{equation}
To obtain the quadratic part of the resonance hamiltonian (\ref{eq:Hpr})
from Eq.~(\ref{eq:Hparpump}) two additional approximations are necessary:
the second line on Eq.~(\ref{eq:Hparpump}) 
involving the combination $\cos ( \omega_0 t ) A_{\bd{k}} a^{\dagger}_{\bd{k}} a_{\bd{k}}$
has to be dropped,
while in the last line one should substitute
 \begin{equation}
 \gamma_{\bd{k}} \cos ( \omega_0 t ) \rightarrow 
 \frac{\gamma_{\bd{k}}}{2} e^{ - i \omega_0 t } \; \; , 
 \; \;
\gamma_{\bd{k}}^{\ast} \cos ( \omega_0 t ) \rightarrow 
 \frac{\gamma_{\bd{k}}^{\ast} }{2} e^{  i \omega_0 t }.
 \label{eq:resonanceapprox})
 \end{equation}
Apparently this approximation has been accepted for many decades in the 
literature\cite{Zakharov70,Tsukernik75,Vinikovetskii79,Lim88,Kalafati89,Lvov94,Rezende09}. However, a thorough study of the non-resonant terms neglected in 
this approximation has been performed by Zvyagin {\it{et al.}}\cite{Zvyagin82}, 
who showed that the neglected terms can qualitatively change the results
obtained in resonance approximation.
Here we would like to point out that the approximations leading to
Eq.~(\ref{eq:resonanceapprox}) amount to 
an essential modification of the original spin hamiltonian. To see this,
let us for the moment accept the validity of these approximations,
thus replacing Eq.~(\ref{eq:Hparpump}) by the non-interacting part
of the resonant hamiltonian (\ref{eq:Hpr}),
\begin{eqnarray}
 \hat{H}_{ 2} ( t ) &  \approx &  
\sum_{\bd{k}} \epsilon_{\bd{k}} a_{\bd{k}}^{\dagger} a_{\bd{k}} 
 \nonumber
 \\
& + &  
\frac{1}{2} \sum_{\bd{k}}
\left[    
\gamma_{\bd{k}} e^{ -i \omega_0 t }  a^{\dagger}_{\bd{k}} a^{\dagger}_{ - \bd{k}}
+ \gamma_{\bd{k}}^{\ast} e^{  i \omega_0 t }  a_{- \bd{k}} a_{\bd{k}}    
  \right],
 \label{eq:H2pr}
 \hspace{10mm}
 \end{eqnarray}
where we have dropped the constant terms.
Using now the inverse of the Bogoliubov transformation (\ref{eq:Bogoliubov}) to re-express
the magnon operators 
in Eq.~(\ref{eq:H2pr})
in terms of  Holstein-Primakoff bosons 
and assuming for simplicity that $\gamma_{\bd{k}}$ is real,
the second term
in Eq.~(\ref{eq:H2pr}) can be written as
 \begin{eqnarray}
 & &
\frac{1}{2} \sum_{\bd{k}}
\left[    
\gamma_{\bd{k}} e^{ -i \omega_0 t }  a^{\dagger}_{\bd{k}} a^{\dagger}_{ - \bd{k}} +
\gamma_{\bd{k}} e^{  i \omega_0 t }  a_{- \bd{k}} a_{\bd{k}}      
 \right]
 \nonumber
 \\
 & =   &\frac{1}{2}   \sum_{\bd{k}} 
 \biggl\{ \frac{ \gamma_{\bd{k}}   A_{\bd{k}}}{\epsilon_{\bd{k}} } 
 \cos ( \omega_0 t ) \left[ 
b^{\dagger}_{\bd{k}} b^{\dagger}_{-\bd{k}} +
b_{ - \bd{k}} b_{\bd{k}}   \right]
 \nonumber
 \\
 & & \hspace{10mm} + i   \gamma_{\bd{k}}  \sin ( \omega_0 t ) 
\left[ 
b^{\dagger}_{\bd{k}} b^{\dagger}_{-\bd{k}} -
b_{ - \bd{k}} b_{\bd{k}}   \right]
\biggr\}
 \nonumber
 \\
 & &   + \sum_{\bd{k}}  \frac{ \gamma_{\bd{k}}  B_{\bd{k}}}{\epsilon_{\bd{k}} } 
 \cos ( \omega_0 t )
 \left[  b^{\dagger}_{\bd{k}} b_{\bd{k}}  + \frac{1}{2} \right].
 \label{eq:terms}
\end{eqnarray}
Only the last term on the right-hand side has the form of the boson representation
of the Zeemann term associated with an external pumping field parallel to
the magnetization, while the first two terms can 
be identified with the boson representation of spin anisotropies
associated with a rotating easy axis perpendicular to the $z$-axis,
see Eq.~(\ref{eq:easyaxis}).
We thus conclude that the time-dependent part of the 
resonant hamiltonian (\ref{eq:Hpr})
does not represent the time-dependent Zeemann energy associated with 
a harmonically oscillating magnetic field
in the direction of the magnetization. Instead, the time-dependent off-diagonal 
pumping terms arise from a rotating easy axis anisotropy perpendicular
to the magnetization. The microscopic origin  of such a term is not clear to us;
possibly
the time-dependent electric field associated with the harmonically varying 
magnetic field parallel to the magnetization can indirectly induce such a term in the spin
hamiltonian, similar to the second order interaction hamiltonian 
in the theory of two-magnon Raman scattering in
antiferromagnets \cite{Fleury68,Elliott69}.
Moreover, in real materials crystallographic or shape anisotropies can
give rise to further contributions to the effective spin hamiltonian which after
Holstein-Primakoff transformation might  
have the same form as the terms in Eq.~(\ref{eq:terms}).

\end{appendix}

\end{document}